# A systematic analysis of biotech startups that went public in the first half of 2021

Sebastian G. Huayamares[1,2,*], Melissa P. Lokugamage[3], Alejandro J. Da Silva Sanchez[2,4,5], James E. Dahlman[1]

[1]Wallace H. Coulter Department of Biomedical Engineering, Georgia Institute of Technology & Emory School of Medicine, Atlanta, GA, 30332, USA

[2]Quantitative & Computational Finance Program, Georgia Institute of Technology, Atlanta, GA, 30332, USA

[3]Alloy Therapeutics, Boston, MA, 02421, USA

[4]Petit Institute for Bioengineering and Biosciences, Georgia Institute of Technology, Atlanta, GA, 30332, USA

[5]Department of Chemical Engineering, Georgia Institute of Technology, Atlanta, GA, 30332, USA

*Correspondence: sebas.hm@gatech.edu

## Abstract

Biotechnologies are being commercialized at historic rates. In 2020, 74 biotech startups went public through an Initial Public Offering (IPO), and 60 went through the IPO process in the first six months of 2021. However, the traits associated with biotech startups obtaining recent IPOs have not been reported. Here we build a database of biotechs that underwent an IPO in the first half of 2021. By analyzing leadership, technological focus, clinical trials, and financing, we found that advanced degrees among the leadership, clinical trials, and intellectual property are important factors for biotech startups. The data also suggest that large private rounds can decrease time-to-IPO and affect post-IPO stock performance. Notably, these traits were often exhibited by the 138 biotech IPOs in 2018-2019, suggesting 2021 data were not driven by COVID.

## Methods & Data Availability

**Data Collection.** All the data analyzed is publicly available. We used Endpoints IPO Tracker[1] to identify Q1 and Q2 2021 IPOs as well as the CEO name, stock ticker, exchange platform, IPO date, amount raised at IPO ($M), and share value at IPO ($). Biotech IPOs that took place in 2018-2019 were identified using NASDAQ's IPO calendar (https://www.nasdaq.com/market-

activity/ipos). We used Clinicaltrials.gov to study active clinical trials, their current phases, locations where the trials were conducted, dates when the trials were first posted, and completions. We used Lens.org, the global patent database recommended for analytics in the USPTO website[2], to analyze patents. We obtained the number of US trademarks filed, their status, and their filing date from the US Patent and Trademark Office databases (Uspto.report). Separately, we compiled information on the technology and therapeutic area of each company. We obtained the year founded, name(s) of the founder(s), city and country of headquarters, valuation at IPO, previous fundraising rounds, the investing team, and location through Crunchbase. We found information on CEO/founder education and previous industry experience using LinkedIn. Stock prices until December 1, 2021, were obtained from Yahoo Finance. All the data are open source: https://bit.ly/32SyoXA.

**Data Processing, Analytics, and Visualization.** Data analytics were performed using Python on Jupyter: https://bit.ly/3mOMXCO. The pie charts and bar plots were generated in Prism9 after analyzing the data in Jupyter. The geographic and bubble maps were plotted using Google Data Studio. The data are also open source: https://datastudio.google.com/s/nRfXzvMy5Tg.

**Introduction**

The average US life expectancy increased from 70 years in 1960 to 79 years in 2019[3], and mortality rates for many diseases are decreasing globally[4]. In addition to non-medical improvements[5, 6] and reduced infant mortality[7], increased life expectancy can be attributed in part to biotechnologies such as vaccines, small-molecule drugs, diagnostics, and devices that enable safer medical procedures[8, 9]. In turn, these biotechnologies are enabled by policies that incentivize innovation and entrepreneurship[10]. However, although an all-time high of 74 biotech initial public offerings (IPOs) occurred in 2020[11] and 60 IPOs took place in the first six months of 2021[1], the traits associated with startups that recently underwent an IPO have been understudied. We therefore analyzed the 60 biotechs that obtained an IPO in Q1 and Q2 of 2021, focusing on the CEO and founder background, technology, clinical trials, and financing. These data suggest that the background of company leadership, clinical trials, intellectual property, amount of private fundraising, and time-to-IPO are important factors for biotech startups when going public. To control for pandemic-related effects, we analyzed the 138 biotech IPOs that took place in 2018-2019 and found similar trends, suggesting these traits may be COVID-independent.

**Biotech CEOs and founders have different backgrounds**

We first analyzed the educational and professional backgrounds of the CEOs leading these 60 biotech startups. Nearly all had a degree beyond a bachelor's (**Fig. 1a**): 37% had a PhD (5% also had an MD or MBA), 26% had an MD (some of whom also had an MBA), and 28% had an MBA. Less frequently represented advanced degrees included JD, PharmD, and MS. Notably, all the biotech CEOs held a college degree, contrasting with software and web-based tech startups[12]. While 27% came from the financial services industry, the majority came from biotech (29%) or pharma (39%) (**Fig. 1b**). While these data suggest that deep technical expertise is important, we also noted that only 3% of the CEOs came from academia. We then compared the backgrounds of CEOs to the backgrounds of the founders. We found that 63% of founders had a PhD and, in many cases, an MD, MBA, or JD in addition to a PhD (**Fig. 1c**). Moreover, unlike with the CEOs, the most common previous experience among founders was a senior position in an academic lab (**Fig. 1d**). Specifically, most founders were principal investigators or recent graduates from the lab where the technology was developed. Consistent with this, many founders retained a technical but non-CEO leadership role such as Chief Scientific Officer or Chief Medical Officer. Finally, we also found that the majority (57%) were first-time founders (**Fig. 1e**). These data led us to three conclusions. First, the market values CEOs with operational or financing experience more than those who developed the original technology. Second, academic labs play an important role in the biotech ecosystem by generating ideas and proof-of-concept data that evolve into startups as well as providing technical leadership. Finally, biotech founders typically possess more formal training than tech founders.

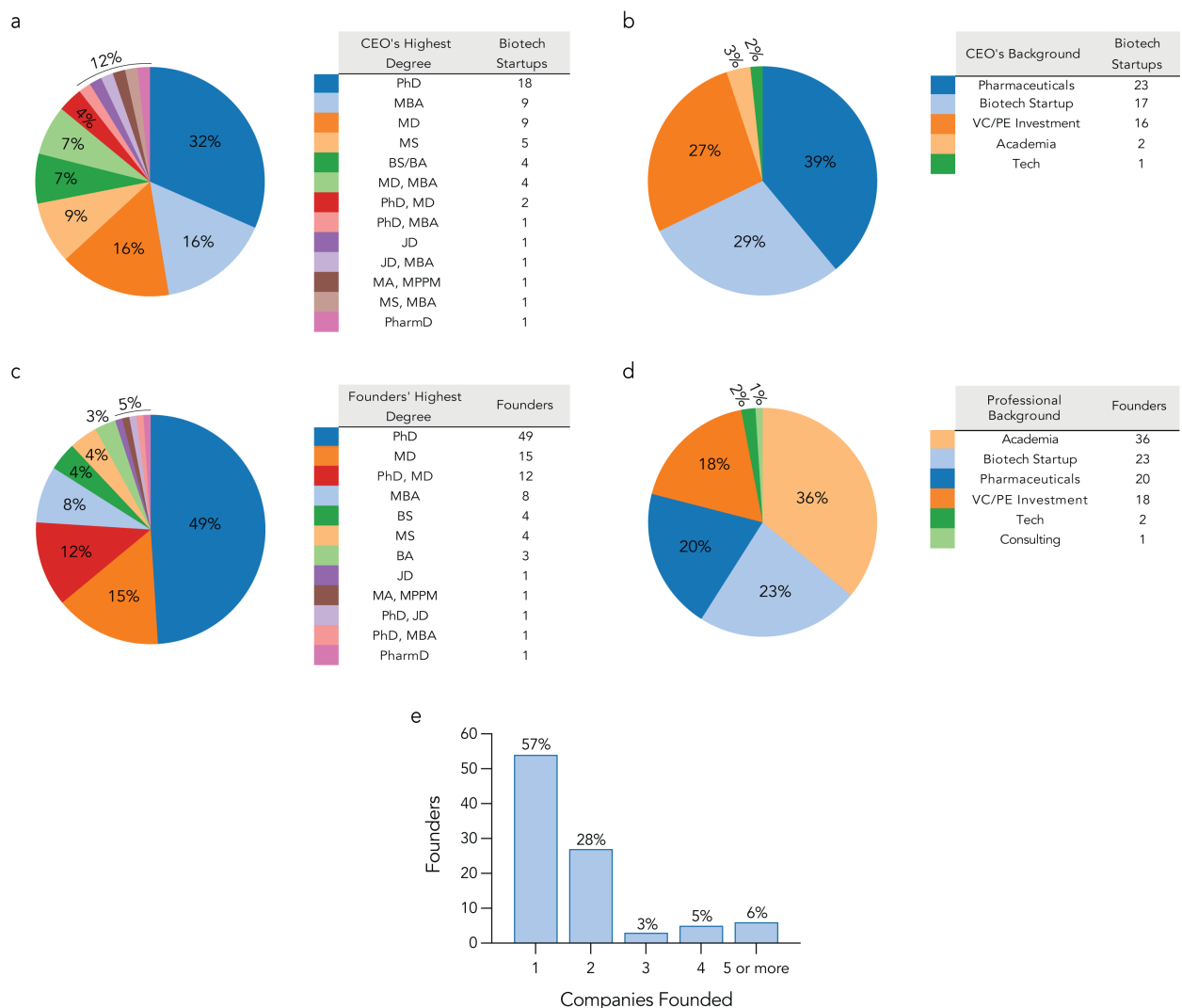

**Figure 1. Chief Executive Officer (CEO) and Founders' Background.** Biotech startups' CEOs classified by **a)** highest degree obtained and **b)** professional experience immediately before assuming the role of CEO at their respective startups. Biotech startups' founders classified by **c)** highest degree obtained, **d)** professional experience immediately before founding their respective startups, and **e)** number of companies founded.

**Biotechs span technologies and therapeutic areas**

We then identified the scientific areas studied by these companies. We noted that 42% specialized in more than one technology or therapeutic area; in these cases, we counted them in all the relevant categories. The most common technology was small molecules, which were developed by 40% of the companies (**Fig. 2a**). The second most common technology was engineered peptides and proteins, which were developed by 18% of the companies. Interestingly, the two bestselling drugs in 2019, which accounted for 33% of the top 10 drugs' total sales, were not small molecules: Humira, a monoclonal antibody targeting tumor necrosis factor, sold $19.6 billion[13], while Keytruda, a monoclonal antibody targeting PD1, sold $11.1 billion[13]. Thus, while small molecules constitute a large share of drugs under development[14], market forces are likely to continue to support biologics[15]. The third most common therapeutic class, represented in 15% of startups, was cell therapies (e.g., CAR-T cells and allogeneic stem cells), while 6% of the startups had expertise in gene therapies. In the future, we expect the number of cell and gene therapy IPOs to increase for several reasons. First, the mRNA-based COVID-19 vaccines developed by Acuitas/Pfizer/BioNTech[16] and Moderna[17] hint at the potential impact of modular nucleic acid-based therapies. For example, Alnylam has developed 13 clinical programs with a single drug delivery system[18] by simply changing the siRNA. Second, it has been estimated that the Food and Drug Administration (FDA) could approve 10 to 20 new gene therapies every year starting in 2025[19]. Third, several clinical datasets suggest that cell therapies can lead to robust therapeutic responses[20-23]. Fourth, we noted that many of the cell therapies focused on oncology, and that oncology was the most common therapeutic area, with nearly threefold more companies focused on cancer than neurological disease, the second-most common disease area (**Fig. 2b**). When compared to the 138 biotech startups that obtained their IPO in 2018-2019, we observed similar technologies and therapeutic areas. Small molecules, protein engineering, and cell and gene therapies were the major technologies under development by these biotechs (**Fig. 2c**), and oncology was also the dominant therapeutic area among these 2018-2019 IPOs (**Fig. 2d**). These 2018-2019 data suggest that the trends identified among the 2021 biotechs were driven by the pandemic. This focus on cancer is likely to continue; it is currently the second leading cause of death in the United States and is projected to be the leading cause of death[24]. Relatedly, given the complexity of manufacturing cell and gene therapies, we anticipate that the number of biomanufacturing IPOs will increase.

Finally, we noted that 6% of the 2021 biotech IPOs had expertise in artificial intelligence and machine learning (AI/ML). The licensing of AI/ML platforms by Sanofi to hunt for metabolic-disease therapies as well as by Genentech to search for cancer treatments[25], along with extensive financing of companies applying AI in drug discovery in 2020-2021[26], demonstrate that AI/ML may be used for target discovery and optimization. In another recent example, Recursion Pharmaceuticals, one of the companies in our dataset, received $150 million in upfront cash from Roche and Genentech, establishing an alliance that could expand to 40 programs[27]. AI is separately projected to be useful in the generation of patents themselves[28]; we therefore anticipate biotech companies using AI to improve their IP portfolios.

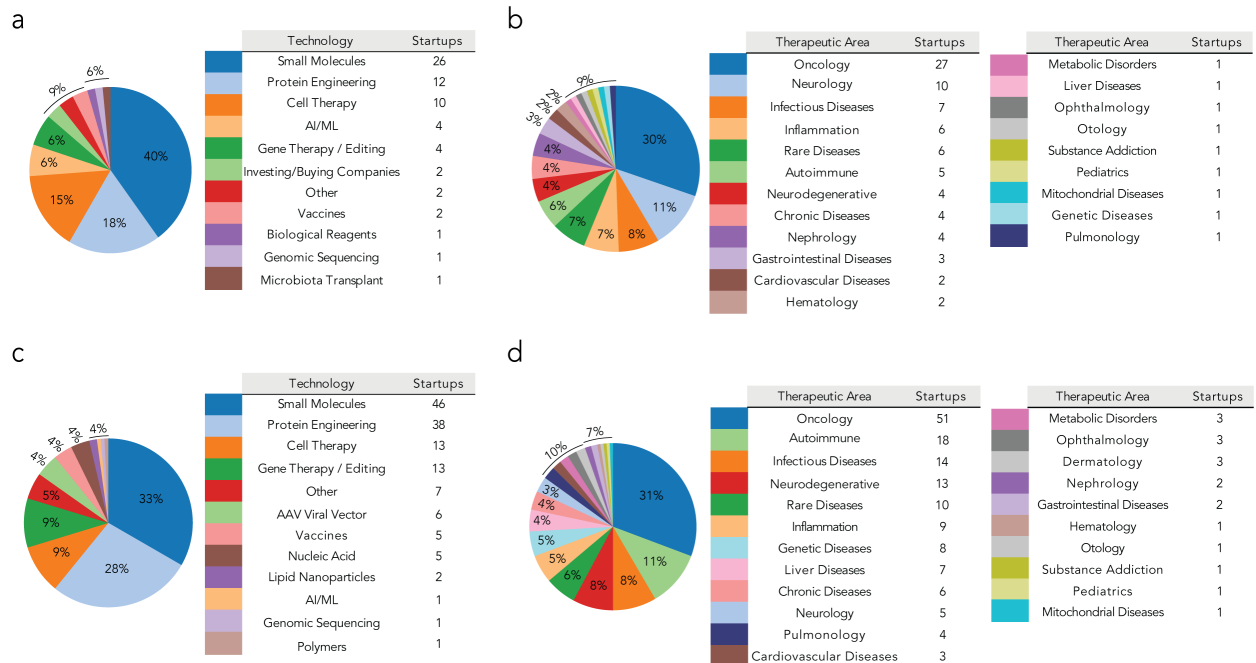

**Figure 2. Biotechnologies and Therapeutic Areas.** Biotech startups that obtained their IPO in 2021 Q1-2 classified by **a)** technology and **b)** therapeutic area. Biotech startups that obtained their IPO in 2018-2019 classified by **c)** technology and **d)** therapeutic area.

### Many companies are at or near clinical stage when they IPO

We reasoned that IPOs could be influenced by the clinical stage of the company. We found that 70% of the companies had at least one clinical trial initiated at the time of IPO (**Fig. 3a**), and a few cases in which companies developing platform technologies initiated multiple clinical trials. For example, Immunocore, which develops immunotherapies and engineered proteins for oncology, autoimmune disorders, and infectious diseases, had six active clinical trials at the time of IPO. Of these companies with clinical trials, the most advanced program that was recruiting was in Phase I (35%), II (48%), or III (18%) (**Fig. 3b**). This suggests that most companies raised private funding sufficient to start clinical trials. Among the 2018-2019 biotechs, only 53% had at least one clinical trial initiated at the time of IPO but over 90% of these trials were at Phase II-III (**Supplementary Fig. 1a-b**). We also noted a strong correlation ($R^2 = 0.98$) between the years taken by the startups to IPO and the years to start recruiting for their most advanced clinical trial (**Fig. 3c**), suggesting that public money raised at IPO is likely being used to further advance their lead clinical programs, which were started prior to their IPO with private funding. While this correlation was not as strong among the 2018-2019 IPOs ($R^2 = 0.70$), we noticed that 70% started at least one clinical trial post-IPO and 53% completed clinical trials post-IPO (**Supplementary Fig. 1c-e**); these data support the hypothesis that the money raised at IPO is used to further their lead clinical programs. Notably, while the United States was the most common site for clinical trials, companies conducted trials in various different countries (**Fig. 3d, Supplementary Fig. 1f**), including countries with no IPO companies during the time we studied (**Supplementary Fig. S2**). We therefore concluded that the location of the startup does not dictate the location of the clinical studies.

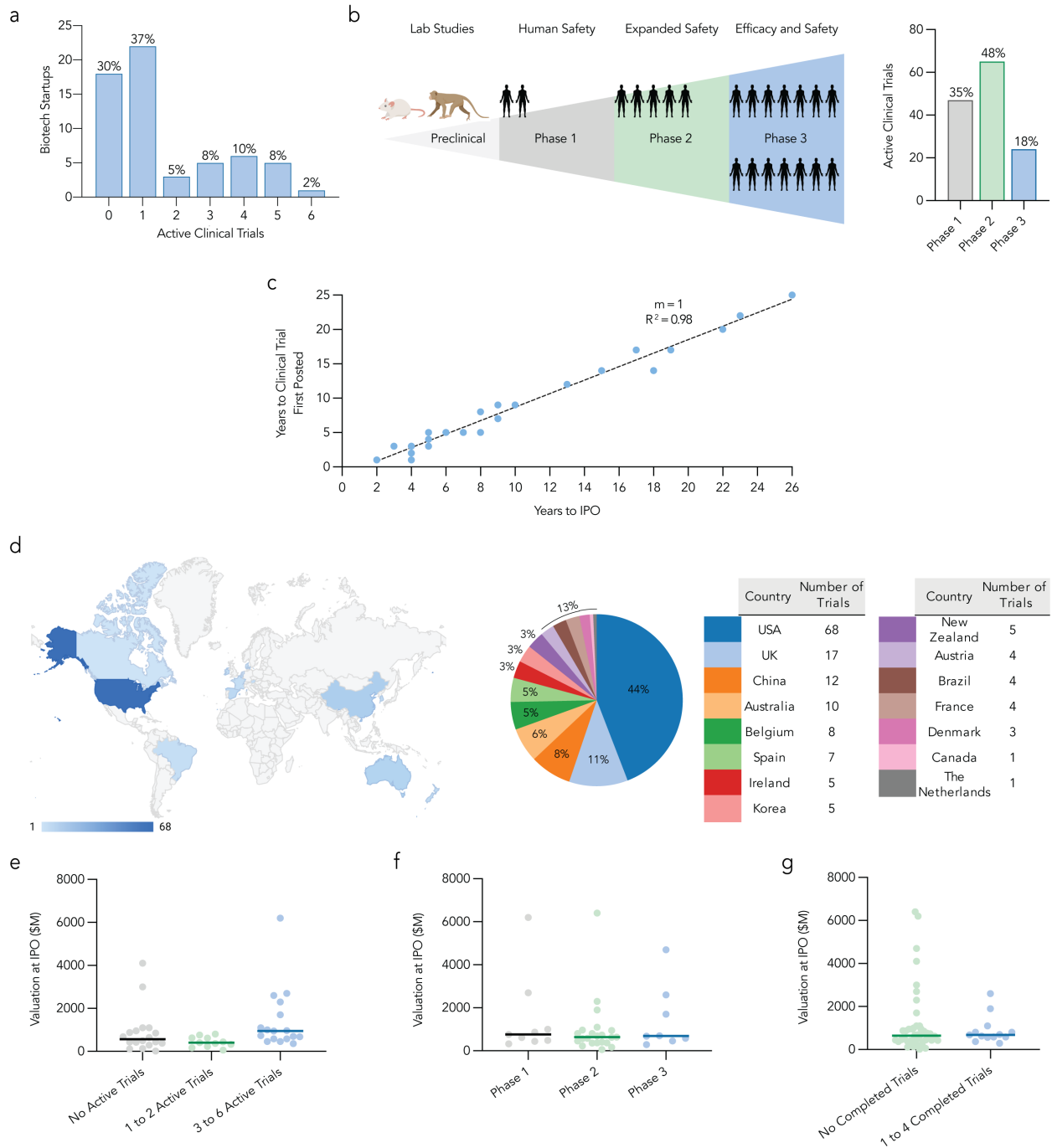

**Figure 3. Clinical Validation. a)** Active clinical trials by the time of IPO. **b)** Phases of the active clinical trials. **c)** Comparison between years taken by the startups to IPO and years taken to post their most advanced clinical trial. **d)** Countries where the biotech startups are conducting their active clinical trials. Valuation of biotech IPOs with **e)** none, 1–2, and 3–6 active clinical trials; **f)** trials in Phase 1, 2, and 3; **g)** none, or 1–4 completed clinical trials by the time of IPO.

Given that most companies had at least one clinical trial at the time of IPO, we wondered whether (i) the number, (ii) the stage, and (iii) the completion of these trials increased valuation at IPO. We found that companies with more active clinical trials had higher valuations on average, although the differences were not significant (**Fig. 3e**). By contrast, the stage of these trials (**Fig. 3f**) and the number of clinical trials completed (**Fig. 3g**) did not increase valuation. These trends were also observed among the 2018-2019 biotechs (**Supplementary Fig. S1g-i**).

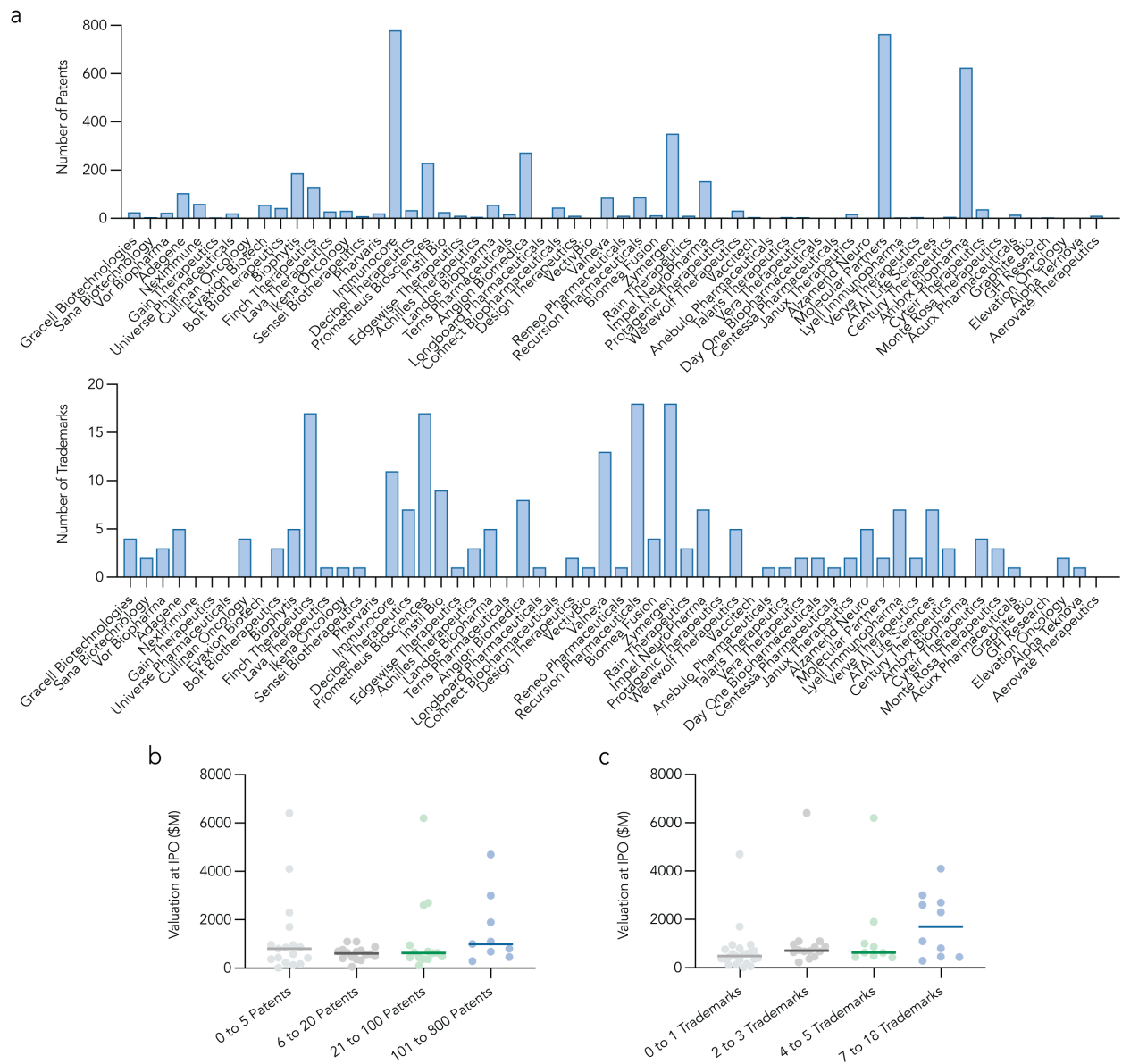

**Figure 4. Intellectual Property. a)** Patents and trademarks filed by the biotechs analyzed. Relationship between valuation at IPO ($M) and **b)** patents and **c)** trademarks filed by the time of IPO.

We also hypothesized that higher valuations could depend on the amount of intellectual property filed by the startups, given the central role that technology licensing plays in the biotech industry[29]. To test this, we analyzed the number of patents and trademarks (**Fig. 4a, Supplementary Table S1**). We first noticed that patents were more common than trademarks, with several companies issuing over 100 patents. Second, while there were large differences in the number of patents filed, we found a more uniform distribution on the number of trademarks owned by the startups. Finally, we found that companies with a higher number of patents (**Fig.**

**4b**) and higher number of trademarks (**Fig. 4c**) tended to have higher valuations. These data suggest that intellectual property can increase valuation.

**Large private rounds can decrease time-to-IPO and increase valuation at IPO**

We also hypothesized that large private rounds could accelerate the time-to-IPO. Notably, nearly half (47%) of the companies took four or fewer years to IPO: 12% of companies reached the public markets one to two years after formation, 35% after three to four years, and 28% after five to nine years, while the remaining 25% required more than 10 years (**Fig. 5a**). We found a similar distribution among the time-to-IPO of the 2018-2019 biotechs (**Fig. 5b**). We calculated the total private fundraising before IPO (**Fig. 5c**) and found that startups that took three or four years to IPO raised the largest sums of money, whereas those that took more than 10 years raised the lowest. Similarly, companies that took three or four years to IPO had the highest valuation (**Fig. 5d**) and raised the largest sums (**Fig. 5e**) at IPO. Notably, 72% of the pre-IPO private money was raised in the United States (**Supplementary Fig. S3**).

We reasoned that the rapid progression from formation to IPO could be driven by early-stage investing economics. We found that biotechs that obtained their IPO within four years tended to only raise a Series A or, alternatively, Series A and B funding (**Fig. 5f-g**). By contrast, biotechs that took more than 10 years to IPO often raised Series A, B, C, and D rounds. Taken together, these data suggest that larger investment rounds accelerate the time to IPO and provide early investors with the opportunity to see gains more quickly and with less dilution.

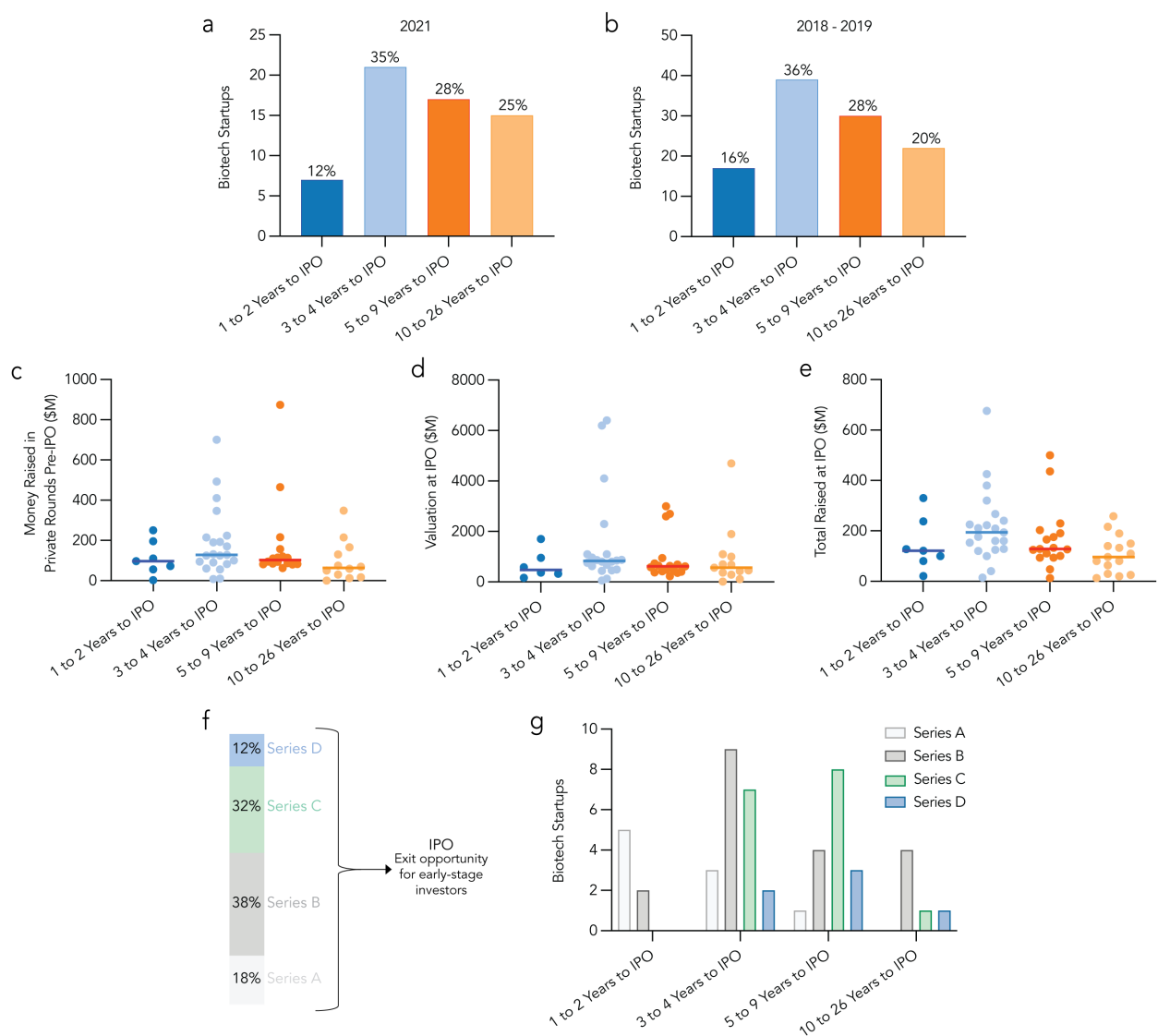

**Figure 5. Pre-IPO Entrepreneurial Finances**. Distribution of years taken for the biotech startups to obtain their IPO (time-to-IPO) used to subdivide them, for the **a)** 2021 IPOs and **b)** 2018-2019 IPOs. **c)** Cumulative money raised by the startups prior to their IPO ($M). **d)** Valuation ($M) of the biotech startups and **e)** money raised ($M) at the time of their IPO. **f)** Stage of all the startups' latest pre-IPO VC rounds, excluding minor debt financing, private equity, and undisclosed rounds. **g)** Last pre-IPO VC round obtained by the startups, subdivided by time-to-IPO.

**Very few new biotech IPOs outperform the S&P 500 index**

After characterizing the companies before IPO, we analyzed their performance after public listing. To control for increased volatility of the stock market in 2021 compared to historical norms[30, 31], we compared the year-to-date stock performance of the 60 companies to the S&P 500 index, from their entrance into the public market until December 1, 2021 (**Fig. 6a**). For example, Vera Therapeutics underwent IPO on May 14; we compared the offer price institutional investors paid to own shares days prior to IPO to the closing price of the last day of our analysis, December 1, 2021, and then compared this to the S&P 500 index from May 14 to December 1. Using these normalized comparisons, we found that 88% of the biotech stocks had a lower price on December 1 compared to their offering price. Notably, only 11% of the 2021 biotechs outperformed the S&P 500 index, suggesting that this was not driven by market volatility. Similarly, we analyzed the stock performance of the 138 biotechs that went public in 2018-2019, from their IPO date to December 1, 2019, and compared it to the S&P 500 index (**Fig. 6a, Supplementary Fig. S4**). Only 28% of the 2018-2019 biotechs outperformed the S&P 500 index, which increased by 16.5% in the two-year period prior to the pandemic. Interestingly, when looking at the stock performances in relation to time-to-IPO, we found similar performance trends between the 2021 IPOs and the 2018-2019 IPOs, suggesting that post-IPO performance of the 2021 biotech IPOs was similar to pre-pandemic IPOs.

We then subdivided the companies based on the time from formation to IPO and tracked their returns normalized to their offering price. All companies that took one or two years to IPO experienced share prices that were flat or lower than their offering price (**Fig. 6b**). By contrast, 24% of the companies that took three to four years to IPO had higher prices than their offering price, and 10% outperformed their normalized S&P 500 index return (**Fig. 6c**). Eighteen percent of the companies taking five to nine years to IPO were priced higher than their initial offering price; of this 18%, all of them outperformed the S&P 500 index return (**Fig. 6d**). Similarly, 20% of the companies that took over 10 years to IPO were priced higher than their initial closing day price, all of which outperformed the S&P 500 index return (**Fig. 6e**). We found similar trends in the historical normalized returns of the 2018-2019 biotech stocks (**Supplementary Fig. S4**). Taken together, these data suggest that large rounds can accelerate IPO; however, once the IPO occurs, there is an increased chance of negative responses from the public market.

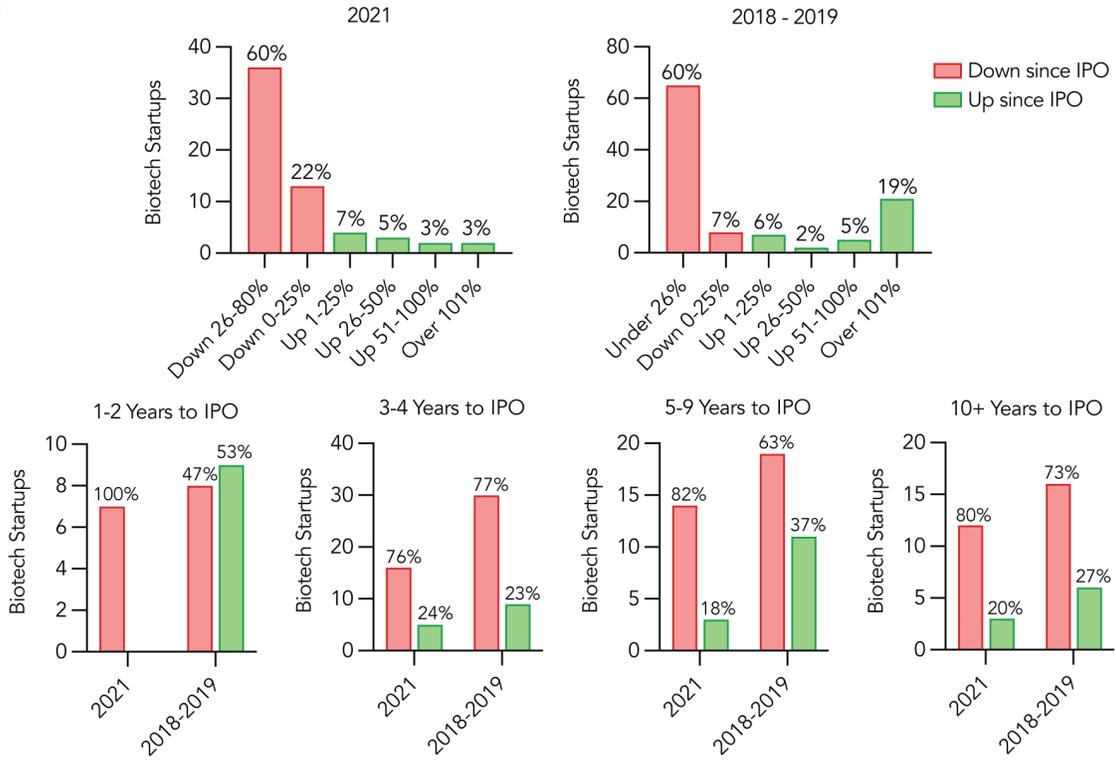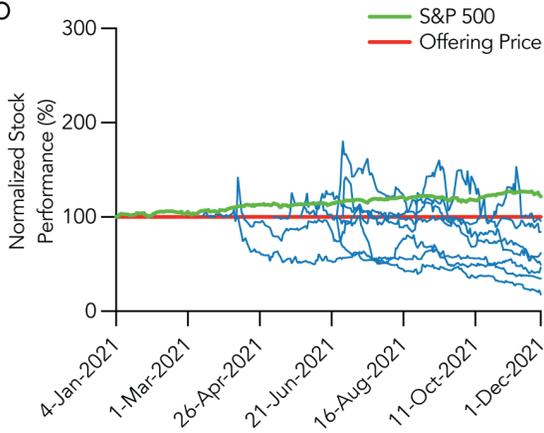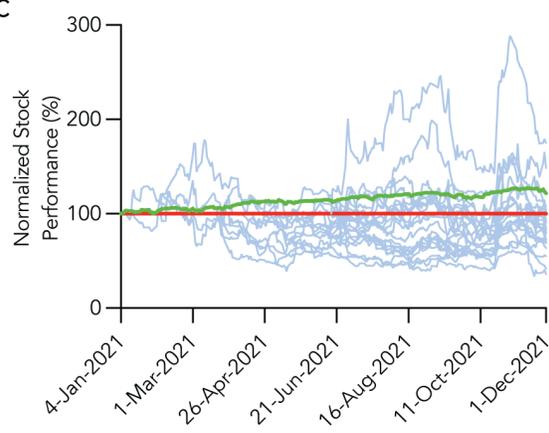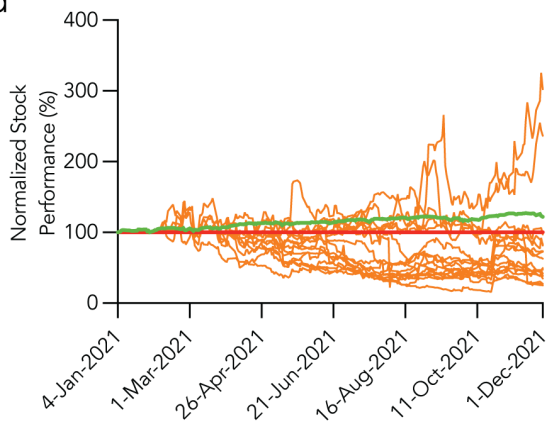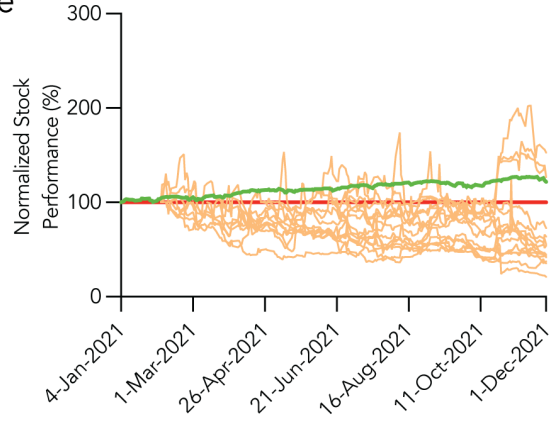

**Figure 6. Post-IPO Stock Performance**. **a)** Stock change from offering price at IPO to December 1, 2021 (for 2021 IPOs) or to December 1, 2019 (for 2018-2019 IPOs); overall and subdivided by time-to-IPO. Historical stock data normalized to their offering price at IPO, compared to the S&P 500 index, for the biotech startups that obtained their IPO in **b)** 1–2, **c)** 3–4, **d)** 5–9, and **e)** 10–26 years.

## Discussion & Conclusions

Biotechs are developing technologies to tackle previously untreatable diseases. However, the need to validate emerging science, optimize practical manufacturing, and overcome clinical risks makes success challenging. By collecting a suite of publicly available datasets, we have identified trends represented in biotechs that overcome enough of these challenges to IPO. These data-driven findings may help entrepreneurs identify traits their companies need to reach the public markets.

Specifically, we found that biotech startups require leaders with industry or business experience as well as scientific founders that understand the technologies they develop. Coupling these teams with established (e.g., small molecules) or emerging technologies with significant potential for growth (e.g., gene therapies) allows companies to raise enough private money to recruit for clinical trials at IPO. Notably, the location of the biotech startup did not limit the location of the clinical trials, which were more distributed than the companies. We also found that many biotech startups went public in four or less years after their inception. Although we do not have direct data to support this, one possible explanation is that early-stage investors may receive quicker returns with less dilution if the company goes public. While this may seem like an advantage, we also observed that stocks did not perform as well as the S&P 500 index. At the same time, 11% of the startups outperformed the S&P 500 return, with one startup (VERA) yielding a 192% return since its IPO. We found similar trends when analyzing the biotech IPOs that took place in the two years prior to the COVID-19 pandemic, suggesting that these findings are not specific to post-pandemic times. Taken together, these data highlight the variability of biotech performances, and suggest that obtaining an IPO is necessary, but not sufficient, for long-term success.

## Acknowledgements

The authors thank Karen Tiegren at Georgia Tech for copyediting the manuscript.

## Author Contributions

Conceptualization: SGH
Methodology: SGH, MPL, and AJDS.
Visualization: SGH, AJDS, MPL, and JED.
First draft: SGH
Draft review and editing: SGH, MPL, AJDS, and JED.

**Competing Interests**

JED is an advisor to GV. MPL is an employee at Alloy Therapeutics. All other authors declare no competing interests.

**Data and Materials Availability**

The database, Python code, and Google Data Studio visualizations are available using the links detailed in the Methods section. All other data are in the main text or supplementary materials.

**Supplementary Information for:**

**A systematic analysis of biotech startups that went public in the first half of 2021**

Sebastian G. Huayamares[1,2,*], Melissa P. Lokugamage[3], Alejandro J. Da Silva Sanchez[2,4,5], James E. Dahlman[1]

[1]Wallace H. Coulter Department of Biomedical Engineering, Georgia Institute of Technology & Emory School of Medicine, Atlanta, GA, 30332, USA

[2]Quantitative & Computational Finance Program, Georgia Institute of Technology, Atlanta, GA, 30332, USA

[3]Alloy Therapeutics, Boston, MA, 02421, USA

[4]Petit Institute for Bioengineering and Biosciences, Georgia Institute of Technology, Atlanta, GA, 30332, USA

[5]Department of Chemical Engineering, Georgia Institute of Technology, Atlanta, GA, 30332, USA

*Correspondence: sebas.hm@gatech.edu

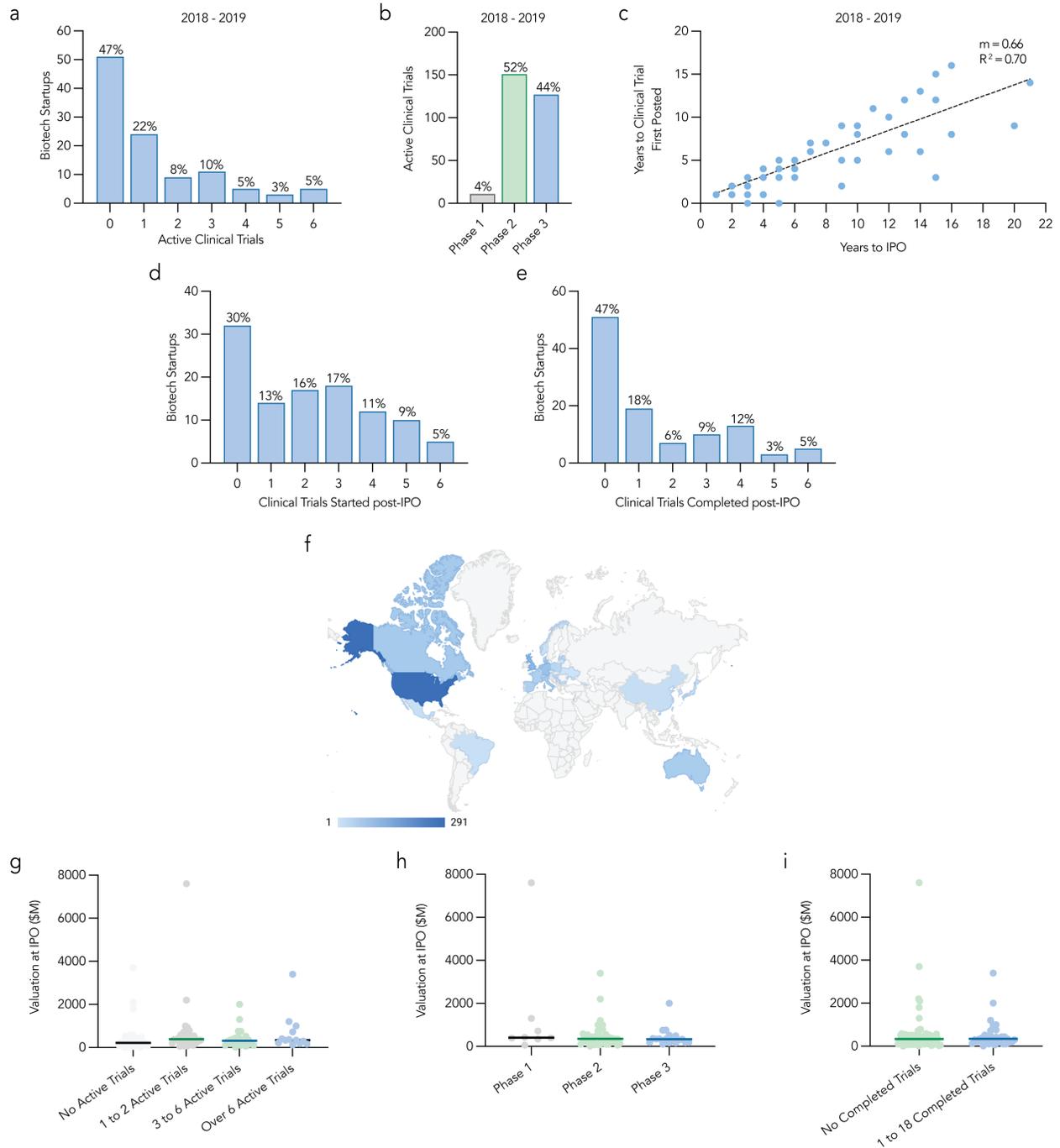

**Supplementary Figure S1. Clinical Validation for 2018-2019 Biotechs. a)** Active clinical trials by the time of IPO. **b)** Phases of the active clinical trials. **c)** Comparison between years taken by the startups to IPO and years taken to post their most advanced clinical trial. Clinical trials **d)** started and **e)** completed post-IPO. **f)** Countries where the biotech startups are conducting their active clinical trials. Valuation of biotech IPOs with **g)** none, 1–2, 3–6, or over 6 active clinical trials; **h)** trials in Phase 1, 2, and 3; **i)** none, or 1–18 completed clinical trials by the time of IPO.

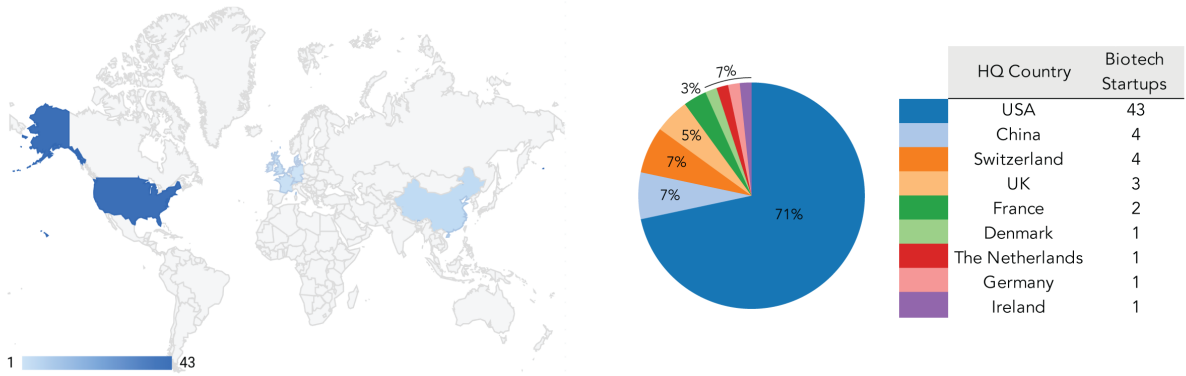

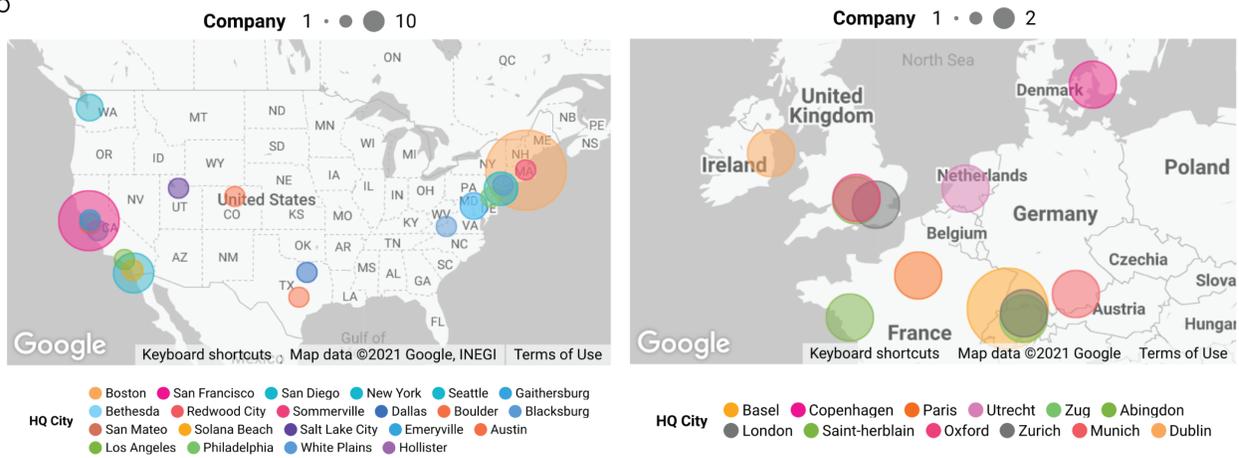

**Supplementary Figure S2. Geographic Distribution of Biotech IPOs. a)** Startup distribution by country of headquarters; **b)** breakdown by city for startups in the U.S. and Europe.

a

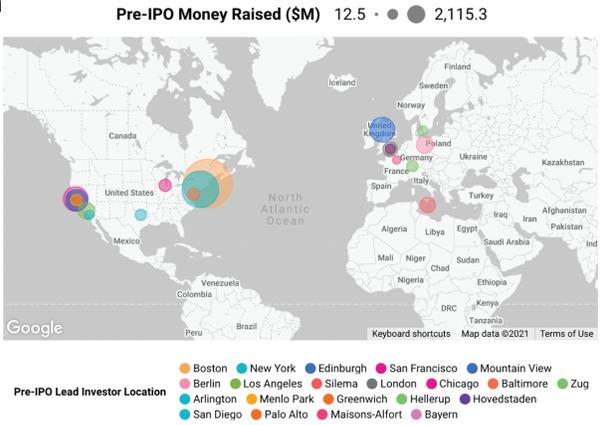

| Investor's Location | Pre-IPO Money Raised ($M) | Lead Investors | Investor's Location | Pre-IPO Money Raised ($M) | Lead Investors |
|---|---|---|---|---|---|
| Boston | $ 2,115.3 | 15 | Zug | $ 170.2 | 1 |
| New York | $ 1,482.2 | 11 | Arlington | $ 166.3 | 1 |
| Edinburgh | $ 874.1 | 1 | Menlo Park | $ 166.1 | 2 |
| San Francisco | $ 866.8 | 6 | Greenwich | $ 95.0 | 1 |
| Mountain View | $ 700.0 | 1 | Hellerup | $ 95.0 | 1 |
| Berlin | $ 465.4 | 1 | Hovedstaden | $ 83.0 | 1 |
| Los Angeles | $ 477.1 | 2 | San Diego | $ 82.4 | 1 |
| Silema | $ 347.1 | 1 | Palo Alto | $ 59.8 | 1 |
| London | $ 329.6 | 2 | Maisons-Alfort | $ 12.5 | 1 |
| Chicago | $ 215.0 | 1 | Bayern | $ 10.0 | 1 |
| Baltimore | $ 195.7 | 1 | | | |

b

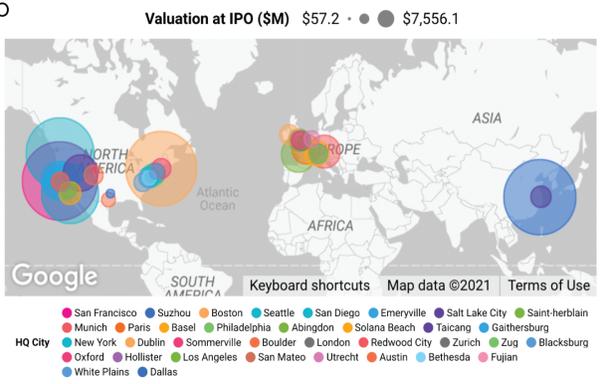
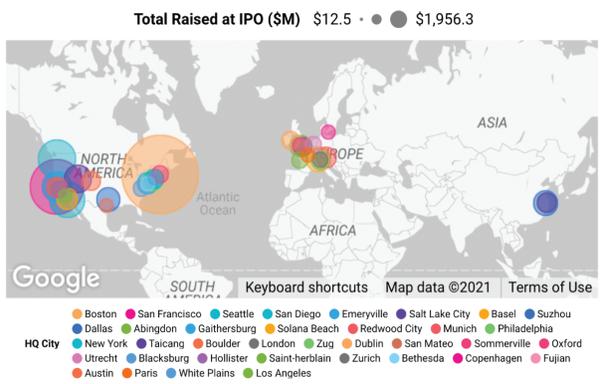

**Supplementary Figure S3. Geographic Trends Behind Investors and Money.** Geographic distribution of **a)** the leading pre-IPO investors quantified by the cumulative amount raised by the startups in private rounds prior to their IPO ($M); and distribution of **b)** the biotech startups, quantified by their valuation and money raised at IPO ($M).

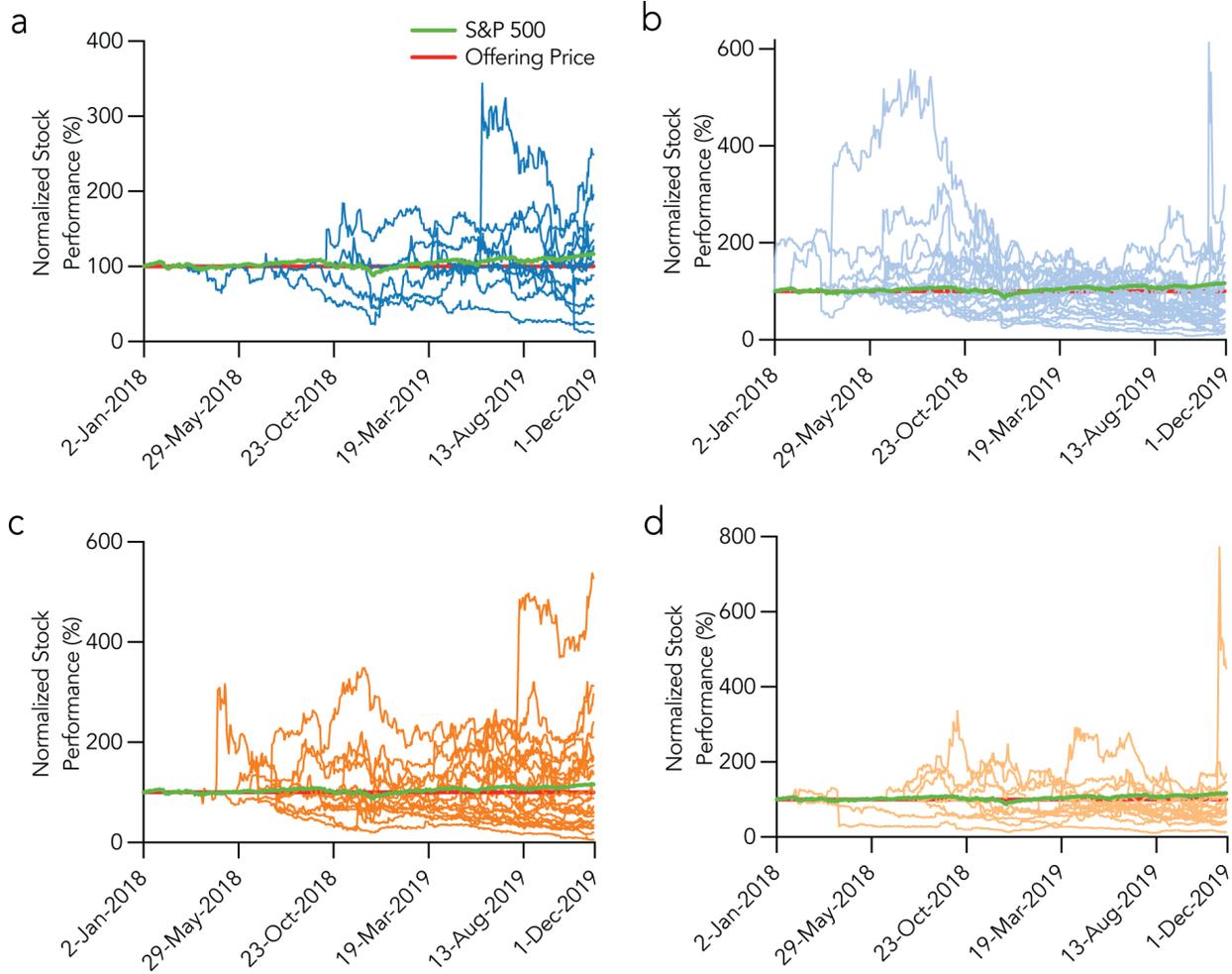

**Supplementary Figure S4. Post-IPO Stock Performance of 2018-2019 Biotechs**. Historical stock data normalized to their offering price at IPO, compared to the S&P 500 index, for the biotech startups that obtained their IPO in **a)** 1–2, **b)** 3–4, **c)** 5–9, and **d)** 10–26 years.

**Table S1.** Intellectual property owned by the biotech startups and their valuation ($M) at IPO.

| Company | Patents Published | US Trademarks | Valuation at IPO ($M) |
|---|---|---|---|
| Immunocore | 780 | 11 | $1,100.0 |
| Molecular Partners | 765 | 2 | $683.3 |
| Ambrx Biopharma | 625 | 0 | $4,700.0 |
| Zymergen | 351 | 18 | $3,000.0 |
| Angion Biomedica | 273 | 8 | $459.5 |
| Prometheus Biosciences | 230 | 17 | $- |
| Biophytis | 187 | 5 | $1,900.0 |
| Impel NeuroPharma | 154 | 7 | $291.2 |
| Finch Therapeutics | 131 | 17 | $801.5 |
| Adagene | 105 | 5 | $1,000.0 |
| Recursion Pharmaceuticals | 88 | 18 | $2,700.0 |
| Valneva | 86 | 13 | $2,600.0 |
| NexImmune | 60 | 0 | $366.4 |
| Evaxion Biotech | 57 | 0 | $- |
| Landos Biopharma | 57 | 5 | $626.6 |
| Connect Biopharmaceuticals | 45 | 0 | $948.9 |
| Bolt Biotherapeutics | 43 | 3 | $690.4 |
| Cyteir Therapeutics | 38 | 4 | $620.4 |
| Decibel Therapeutics | 34 | 7 | $437.4 |
| Werewolf Therapeutics | 33 | 5 | $440.6 |
| Ikena Oncology | 32 | 1 | $481.9 |
| Lava Therapeutics | 29 | 1 | $380.3 |
| Instil Bio | 26 | 9 | $- |
| Gracell Biotechnologies | 25 | 4 | $6,200.0 |
| Vor Biopharma | 24 | 3 | $641.6 |
| Universe Pharmaceuticals | 21 | 0 | $108.8 |
| Pharvaris | 20 | 0 | $636.8 |
| Janux Therapeutics | 18 | 2 | $678.3 |
| Terns Pharmaceuticals | 17 | 0 | $407.5 |
| Acurx Pharmaceuticals | 16 | 1 | $57.2 |
| Biomea Fusion | 13 | 4 | $489.0 |
| Design Therapeutics | 11 | 2 | $1,100.0 |
| Reneo Pharmaceuticals | 11 | 1 | $363.2 |
| Edgewise Therapeutics | 10 | 1 | $761.1 |
| Rain Therapeutics | 10 | 3 | $460.2 |
| Aerovate Therapeutics | 10 | 0 | $323.5 |
| Sensei Biotherapeutics | 9 | 1 | $561.6 |
| Achilles Therapeutics | 7 | 3 | $731.2 |
| Century Therapeutics | 7 | 3 | $1,100.0 |
| Vaccitech | 6 | 0 | $579.1 |
| Talaris Therapeutics | 6 | 1 | $701.0 |
| Verve Therapeutics | 6 | 2 | $876.1 |
| Sana Biotechnology | 5 | 2 | $6,400.0 |
| Vera Therapeutics | 5 | 2 | $225.3 |
| Gain Therapeutics | 4 | 0 | $129.0 |
| Lyell Immunopharma | 4 | 7 | $4,100.0 |
| GH Research | 4 | 0 | $808.3 |
| Protagenic Therapeutics | 3 | 0 | $17.5 |
| Graphite Bio | 3 | 0 | $951.6 |
| Cullinan Oncology | 1 | 4 | $869.3 |
| Anebulo Pharmaceuticals | 1 | 1 | $162.9 |
| Monte Rosa Therapeutics | 1 | 3 | $845.3 |
| Elevation Oncology | 1 | 2 | $365.2 |
| Longboard Pharmaceuticals | 0 | 1 | $- |
| Day One Biopharmaceuticals | 0 | 2 | $966.8 |
| Centessa Pharmaceuticals | 0 | 1 | $1,700.0 |
| Alzamend Neuro | 0 | 5 | $424.6 |
| ATAI Life Sciences | 0 | 7 | $2,300.0 |
| Alpha Teknova | 0 | 1 | $433.8 |
| VectivBio | 0 | 1 | $578.2 |